\documentclass[12pt]{article}

\usepackage{amssymb}

\usepackage{amsfonts}
\usepackage{latexsym}
\usepackage{amsmath}
\usepackage{amsthm}
\usepackage{geometry} 
\usepackage{graphicx}
\usepackage{verbatim}
\usepackage[colorlinks=true]{hyperref} 
\usepackage{bm} 

\usepackage{color}

\newtheorem{thm}{Theorem}[section]
\newtheorem{prop}[thm]{Proposition}

\newtheorem{remark}[thm]{Remark}

\newtheorem{example}[thm]{Example}
\newtheorem{defi}[thm]{Definition}

\def\lddots{\mathinner{\mkern1mu\raise1pt\hbox{.}\mkern2mu  
\raise4pt\hbox{.}\mkern2mu\raise7pt\vbox{\kern7pt\hbox{.}}\mkern1mu}}
\makeatletter
\def\numberbysection{\@addtoreset{equation}{section}
 \def\theequation{\thesection.\arabic{equation}}}
\makeatother  
  
\numberbysection  

\def\cA{{\cal A}}

    \def\cK{{\cal K}}    
        
        \def\cR{{\cal R}}

\def\fa{{\mathfrak a}}

\def\fr{{\mathfrak r}}

\newcommand{\nonu}{\nonumber\\}  
\newcommand{\be}{\begin{eqnarray}}  
\newcommand{\ee}{\end{eqnarray}} 
\newcommand{\beano}{\begin{eqnarray*}}  
\newcommand{\eeano}{\end{eqnarray*}}

\def\bb{\mathbb}
\def\C{\bb C}

\def\cm{\mathcal}
\def\cR{\cm R}
\def\cA{\cm A}
\def\cK{\cm K}

\newcommand{\CC}{{\mathbb C}}

\newcommand{\II}{{\mathbb I}}

\newcommand{\MM}{{\mathbb M}}

\def\tr{\mathrm {Tr}}

\def\Rep{\mathrm {Rep}}
\def\ldb{\mathopen{\{\!\!\{}} \def\rdb{\mathclose{\}\!\!\}}}

\newcommand{\mb}[1]{\quad\mbox{#1}\quad}
\newcommand{\wt}[1]{\widetilde{#1}}

\newcommand{\sun}{{\scriptscriptstyle \rm I}}
\newcommand{\sdx}{{\scriptscriptstyle \rm I\hspace{-1pt}I}}
\newcommand{\str}{{\scriptscriptstyle \rm I\hspace{-1pt}I\hspace{-1pt}I}}
\newcommand{\atopn}[2]{\genfrac{}{}{0pt}{}{#1}{#2}}

 \title{{\bf Quantization and dynamisation of Trace-Poisson brackets}}
\author{ \textsf{Jean Avan$^a$,
Eric Ragoucy$^b$ and Vladimir Rubtsov$^c$}
\\\\
\textit{$^{a}$ Laboratoire de Physique Th\'eorique et Mod\'elisation (CNRS UMR 8089),} \\
\textit{Universit\'e de Cergy-Pontoise, F-95302 Cergy-Pontoise, France} \\
\\
\textit{$^{b}$ LAPTh, CNRS and Universit\'e de Savoie,} \\
\textit{9 Chemin de Bellevue, BP 110, F-74941 Annecy le Vieux Cedex} \\
\\
\textit{$^c$ Laboratoire  Angevin de REcherche en MAth\'ematiques (CNRS UMR 6093),} \\
\textit{UNAM et Universit\'e d'Angers,  Facult\'e des Sciences,}\\ 
\textit{2, Boulevard Lavoisier, 49045 Angers Cedex and} \\
\textit{Theory Division, ITEP, 25, Bol. Tcheremushkinskaya, }\\
\textit{117259, Moscow, Russia}}
\date{}

\footnotetext[1]{Emails: {\tt avan@u-cergy.fr}, 
{\tt ragoucy@lapth.cnrs.fr},
{\tt volodya@univ-angers.fr}}

\begin{document}

\maketitle
\thispagestyle{empty}
\vfill
\abstract{The quantization problem for the trace-bracket algebra, derived from double Poisson brackets, is discussed.  We obtain a generalization of the 
boundary YBE (or so-called ABCD-algebra) for the quantization of quadratic trace-brackets.
A dynamical deformation is proposed
on the lines of Gervais-Neveu-Felder dynamical quantum algebras.}

\vfill

\rightline{LAPTH-007/14\qquad}
\vfill
\clearpage 
\newpage

\section{Introduction}

There has been a long-standing interest of both communities -- mathematicians and physicists -- in a class of objects known as \textit{character varieties} or \textit{spaces of representations}, and their
equivalence  classes  -- or "moduli spaces". A typical example of such objects is the moduli space ${\mathcal M}_{\Sigma,G}$ of flat connections in a principal bundle with the structure group $G$ over a Riemann
surface $\Sigma$.
This space is defined as a quotient ${\mathcal M}_{\Sigma,G} = {\rm Hom (\pi_1 (\Sigma), G)/G}$ of the representation space of the fundamental group $\pi_1(\Sigma)$ of the surface $\Sigma$ in $G$ with respect to conjugation by the group $G$.

There are many applications of the character varieties from a
geometric viewpoint. The moduli spaces of representations can be used also for describing the moduli of stable
vector bundles over a compact Riemann surface of genus $>2.$ This is the set of equivalence classes of irreducible unitary representations of the fundamental
group. Another classical example is the Teichm\"{u}ller space of Riemann surface complex structures - the set of equivalence classes
of irreducible representations of the fundamental group in $PSL_2(\mathbb R)$ or $SL_2(\mathbb R).$

From a physical point of view, such moduli spaces (or their reductions, subspaces  etc.) act as arena of various scenarii of modern classical and quantum field theories - (super)string theory, (super)gravity and tentatives to unify them. Many of them can be considered as phase spaces of interesting integrable systems (Beauville, Hitchin et al.).

An algebraic avatar of the character varieties is provided by the representation space ${\rm Rep}_N(A)$ 
or more precisely the affine scheme  
$\mathbb C[{\rm Rep}_N(A)]^{GL_N(\mathbb C)}$. Indeed
in an algebraic context the representation spaces of a (non)commutative associative algebra $A$ become algebraic varieties which "approximate", by Kontsevich philosophy, the underlying variety - the spectrum
of the commutative counterpart of $A.$. In particular, if $A$ is a free associative algebra over $\mathbb C$, that is, $A = \mathbb C<x_1,...,x_n>$ is given by a finite number of generators  $x_1,\ldots,x_n$, the
space of $N-$dimensional representations is 
$${\rm Rep}_N(A) = \lbrace M=(M_1,\ldots, M_n)\in {\mathbb M}_N(\mathbb C)\oplus...\oplus {\mathbb M}_N(\mathbb C)\rbrace,$$
where ${\mathbb M}_N(\mathbb C) = {\rm End}(\mathbb C^N).$

The group $GL_N(\mathbb C)$ acts on ${\rm Rep}_N(A)$ by conjugations : $g.M =\left(gM_1 g^{-1},..., gM_n g^{-1}\right)$ and the quotient space ${\rm Rep}_N(A)^{GL_N(\mathbb C)}$ describes 
isomorphism classes of semisimple representations of $A.$ The classical Procesi theorem on $GL_N-$invariants 
\cite{Proc} says that the coordinate ring $\mathbb C[{\rm Rep}_N(A)]^{Gl_N(\mathbb C)}$ is 
generated by traces of "words" in generic matrices $M_1,\ldots, M_n$ (see Example 2.4 below).

A typical geometric character variety carries a Poisson or symplectic structure studied in many classical works by e.g. Atiyah-Bott, Goldman, Hitchin, Beauville, Mukai et al.
Similarly in the above described algebraic context the affine scheme  
$\mathbb C[{\rm Rep}_N(A)]^{GL_N(\mathbb C)}$ can also be supplied with a natural Poisson algebra structure. Recently the 
wonderful algebraic mechanism governing the Poisson algebra structure on the representation spaces was discovered by M. Van den Bergh \cite{VdB1} and in a slightly different but closely related way by  W. Crawley-Boevey \cite{CB}.

The Van den Bergh construction is based on the notion of \textit{double Lie bracket} on an associative algebra $A$ which is a $\C-$bilinear operation $\ldb -,- \rdb : A\otimes A \to A\otimes A$ which satisfies
certain \textit{twisted} Lie algebra axioms. One may add a Leibniz-like property, different for left-hand and right-hand arguments in the correspondence with outer and inner $A-$bimodule structures on
$A\otimes A$. Such an operation is then called a \textit{double Poisson structure} on $A.$ Precise definitions are given in subsection \ref{sect:doublePB}.

A natural composition of a double Poisson bracket with the associative algebra multiplication $\mu$ defines an interesting operation on $A$ and on the space of "traces" on $A$ (the vector space $A/[A,A]$).  This trace space $A/[A,A]$ is identified with the Hochschild homology of degree 0, $HH_0 (A,A)$,  and W. Crawley-Boevey had showed that the projection of  $\mu\circ\ldb -,-\rdb$ defines a Lie algebra structure 
$<-,->$ on $HH_0(A,A)$. This structure then defines a unique Poisson bracket on the invariant representation space $\C[{\rm Rep}(A)]^{GL_N(\mathbb C)}.$ The latter is identified with the trace map ${\rm Tr}: A/[A,A] \to \C[{\rm Rep}(A)]$ image. These Poisson brackets realize a "trace-covariant" object in the following sense:
$$\lbrace{\rm Tr}(a),{\rm T}(b)\rbrace = {\rm Tr}<\pi(a), \pi(b)>\,,$$ 
where $\pi : A \to A/[A,A]$ is the natural projection.

Such "trace-invariant" Poisson structures coming from $A/[A,A]$ appeared in \cite{MikSok} and \cite{OlSok}  in 
the context of some integrable ordinary differential equations on associative algebras. Such brackets were constructed, their Hamiltonian nature in the proper framework of the Hamiltonian formalism was studied and examples of interesting non-commutative models on associative algebras
were given. These works initiated the recent developments, including the formulation of analogues of Yang -Baxter conditions and their relevance to Hamiltonian formalism, together with related Poisson structures in theories of integrable systems with {\it matrix} variables \cite{ORS1}. The special role of the Associative Yang-Baxter Equation and  (anti-) Frobenius algebra structures were uncovered.

In the case of free algebra the related notion of ``trace brackets'' was developed and studied extensively in \cite{ORS2} especially in the case of "quadratic" brackets. In the case of quiver path algebras the trace Poisson brackets on the
moduli space of finite-dimensional representations are related to the pre-hierarchy  of Hamiltonian
structures of Ruijsenaar Schneider models \cite{Biel}.
On the same line, Massuyeau and Turaev \cite{TM1} constructed a graded Poisson algebra structure on representation algebras associated with the loop algebra of any smooth oriented manifold $M$ with non-empty boundary. When $M$  is a Riemann surface $\Sigma$, the corresponding bracket coincides with  the quasi-Poisson bracket on the representation space ${\rm Hom}(\pi_1 (\Sigma), GL_N)$ defined in their previous work \cite{TM2} via the described (slightly more general) algebraic construction of Van den Bergh.

Further studies of \textit{parameter-dependent} AYBE and Poisson structures were done in \cite{ORS3}. It is amusing to observe that the parameter- dependent AYBE were object of interest and studies even earlier
than the "constant" AYBE. They appeared in different contexts: as associativity conditions in the quadratic parameter dependent algebras related to Sklyanin elliptic algebras of Odesskii and Feigin \cite{OdFeig} and as a triple Massey product expression for the associativity constraint in $A_{\infty}-$ category which is the derived category of coherent sheaves on an elliptic curve \cite{Polis}. The solutions of the AYBE relate to triple Massey products for simple vector bundles on elliptic curves and their degenerations. Further study of this connection  has been continued in \cite{BurKreus}.

All these studies have been yet conducted purely on \textit{classical} associative and trace algebras. A natural question is thus
to construct a quantum version of these algebras. A second related question is to adress the issue of consistent deformations of these classical and/or quantum algebra structures. The quantum case is much easier to handle since the underlying algebraic notions of deforming bialgebras are much simpler than the classical ones related to Poisson algebras. We shall tackle these two issues here, 
however restricting ourselves for the time being to \textit{non-parametric} associative and trace algebras.

Let us briefly summarize the content of our work.
We shall first describe the relevant structures, detailing in particular double-brackets (classical) and their subsequent
trace-Poisson algebras. We shall
then give a full $r$-matrix type description of a `parameter-independent'' trace-Poisson algebra, taking the form of a generalized 
$a,s$ Poisson structure \`a la Maillet \cite{Mail}. We shall define its extension to the most general quadratic form, which lead us to a consistent
quantum algebraic structure mimicking the Freidel--Maillet \cite{FM} quadratic exchange algebras, albeit with a new, third vector index. The bivector formulation introduced by Freidel and Maillet plays a crucial role here. By newly deriving the bivector formulation for 
the known three dynamical reflection algebras (without the extra vector index or ``flavor index'') we identify several key features
of a consistent dynamical deformation of quadratic exchange algebras, which we then extend in the most natural way to define a dynamical deformation of the quantum trace reflection algebra.

\section{Double brackets on algebras and their representations}
We introduce here the general, purely algebraic notions of Double Lie and Double Poisson brackets together with the
associated general notion of Trace Poisson brackets to be used extensively in the following. Let us first establish the
algebraic framework.

\subsection{Algebraic generalities }

We suppose that $A$ is an associative finite dimensional $\C-$algebra (with unity).
We will consider $A\otimes A$ as an $A-A$ bimodule with respect to the outer
 and the inner structures, defined respectively by
\be
&&a.(\alpha\otimes\beta).b = (a\alpha)\otimes(\beta b)\,,\qquad \alpha\otimes\beta\in A\otimes A \mb{and} a,b\in A\\
&&a(\alpha\otimes \beta)b = (\alpha b)\otimes(a\beta). 
\ee
The $A\otimes A$-valued derivations $D(A,A\otimes A)$ of $A$ (wrt the outer bimodule structure) will play the role of the usual derivations ("vector fields") on $A$. 

\begin{example} Let $A$ be the free $\C-$algebra with $n$ generators $A= {\C}<x_1,\ldots, x_n>.$
The partial double derivations are defined for each generator $x_{\alpha}$ as
$\partial_{\alpha} \in D(A,A\otimes A)$ such that $\partial_{\alpha}(x_{\alpha})=1\otimes 1$
and $\partial_{\alpha}(x_{\beta}) = 0,\ \alpha\neq \beta,\quad 1\leq \alpha \leq m.$
\end{example}

The $A-$bimodule $\Omega_A$
of 1-differentials is generated by $da$,  $a\in A$, with relations $d(ab) = a(db) + (da)b$ for $a,b\in A$
and we assume $d^2 a= d(da)=0$ and $d(a)=da.$ We have
$$D(A,A\otimes A) = {\rm Hom}_{A\otimes A^{o}}(\Omega_A, A\otimes A).$$
Here $A^{o}$ denotes the "opposite" algebra: if $\mu : A\otimes A \to  A$ is the multiplication 
law in $A: \mu(a\otimes b) = ab$, then $\mu^{o} : A^{o}\otimes A^{o} \to  A^{o}$ is
$\mu^{o}(a\otimes b) = ba.$ We shall denote by $A^{e}:= A\otimes A^{o}$ the enveloping algebra of $A$.

\subsection{Double Lie and Poisson brackets\label{sect:doublePB}}

\begin{defi}\label{DLB} Let $V$ be any $\C-$vector space. A {\bf double Lie bracket} is a 
$\C-$linear map $\ldb,\rdb : V \otimes V \mapsto V \otimes V$ satisfying the following two conditions:
\be\label{skew}
\ldb u, v\rdb = - \ldb v,u\rdb ^{o},
\ee
and
\be\label{db}
\ldb u,\ldb v,w \rdb\rdb_l + \sigma\big(\ldb u,\ldb v,w \rdb\rdb_l\big) +\sigma^2\big(\ldb u,\ldb v,w \rdb\rdb_l\big) =0,
\ee
where $\sigma \in S_3$ is the cyclic permutation $\left(\begin{array}{ccc} 1 &2 &3\\ 2&3&1\end{array}\right)$, whose action is 
defined as follows:

 for $v= v_1 \otimes v_2 \otimes v_3 \in V_1\otimes V_2 \otimes V_3$ 
one has $\sigma(v) = v_{\sigma^{-1}(1)}\otimes v_{\sigma^{-1}(2)} \otimes v_{\sigma^{-1}(3)}$.

 The bracket in \eqref{db}, $\ldb u,\ldb v,w \rdb\rdb_l$ has to be understood as an extension of the operation 
$\ldb -,-\rdb\in {\rm End}(V \otimes V)$ by
$$\ldb u, v\otimes w \rdb_l :=\ldb u,v\rdb \otimes  w$$ 
and defines an element in $V\otimes V\otimes V$. 
\end{defi}

If we replace the vector space $V$ by an associative $\C-$algebra $A$ we must examine the compatibility of
the double Lie bracket \eqref{DLB} with the associative multiplication.
We suppose, following M. Van den Bergh \cite{VdB1}, that the double Lie bracket satisfies
the "usual" Leibniz rule on the right argument:
\be\label{RLR}
\ldb a, bc \rdb = b\, \ldb a,c \rdb  + \ldb a,b \rdb\, c\,,
\ee
for any $a,b,c\in A$. In other words  the operation $\ldb a,-\rdb : A\to A\otimes A$ is a double derivation
with values in the bimodule $A\otimes A$ with the outer $A-$bimodule structure.
The derivation property with respect to the left argument follows from the skew-symmetry of the
double Lie bracket and from \eqref{RLR}:
\be\label{LLR}
\ldb ab,c \rdb = a\, \ldb b,c \rdb  + \ldb a,c \rdb\, b\,.
\ee
This means that the operation  $\ldb -,b\rdb : A\to A\otimes A$ is a double derivation
with values in the bimodule $A\otimes A$ with the inner $A-$bimodule structure.
\begin{defi}\label{DPB}
A double Lie bracket $\ldb-,-\rdb : A\times A \to A\otimes A$ on an associative $\C-$algebra $A$
 satisfying  \eqref{RLR} is called a {\bf double Poisson bracket on $A$}.
\end{defi}

\subsection{From double Poisson brackets to trace Poisson brackets}

\subsubsection{The affine varieties of representations and their coordinates}
Following  \cite{VdB1}, \cite{ORS1} and \cite{TM2} we now define classes of 
Poisson structures on representation spaces of the given associative algebra $A$.  For any natural
number $N \geq 1$ we define an algebra $A_N$ whose commutative generators are defined by a correspondence from
elements $a\in A: a\to a_{ij}, 1\leq i,j\leq N$ which satisfy the standard $\C-$matrix element relations:
$$(a+b)_{ij} = a_{ij}+b_{ij},\quad (ab)_{ij}= \sum_k a_{ik}b_{kj},\quad 1_{ij}=\delta_{ij},$$
where $a,b\in A$ and $1\leq i,j\leq N.$
In other words, there is a canonical bijection between $A_{N}^* = {\rm Hom}_{\C}(A_{N},\C)$ and ${\rm Hom}(A, {\mathbb M}_N(\C))$
which assign to each linear functional $l\in A_{N}^*$ the algebra morphism ${\hat l}:A\to {\mathbb M}_N(\C)$ defined
by $[{\hat l}(a)]_{ij} = l(a_{ij})$. Reciprocally if we have a map $l:A \to {\mathbb M}_N(\C)$ then it corresponds by the 
bijection to the linear map $A_N \to \C$ such that the generator $a_{ij}\in A_N$ is transformed in $i,j-$component of the matrix $l(a).$
Geometrically, one can say that there exists an affine variety $\Rep_N (A)$ whose coordinate algebra $\C[\Rep_N (A)]$ is the
commutative algebra $A_N.$
\subsubsection{Trace map}
The usual matrix trace defines correctly the map $\tr : A\to A_N$ by $\tr(a) := \sum_i a_{ii}$ for any $a\in A$. This
map annihilates the commutators : $\tr(ab-ba)=0$ for any $a,b\in A$ and extends to the map
$\tr : A_{\natural} \to A_N$ where $A_{\natural}:= A/[A,A]$ is the quotient vector space (the "trace
space of $A$").
We shall denote in what follows the projection of $a\in A$ in $A_{\natural}$ by $p(a).$

\begin{example}\label{free} 
Consider the case of the free associative algebra
$A=\C<x_1,\ldots,x_m>.$ 
 
The coordinate algebra $\C[\Rep_N(A)]$ in this case is the polynomial ring of $mN^2$
variables $x^j_{i,\alpha},$ where
$$x_{\alpha} \to M_{\alpha}=\left( \begin{array}{ccc}
x^1_{1,\alpha}& \cdot& x^N_{1,\alpha}\\
\cdot& \cdot& \cdot\\
x^1_{N,\alpha}& \cdot& x^N_{N,\alpha}
\end{array}\right), \qquad  1\leq \alpha\leq m. $$

The map ${\tr}$ gives the following interpretation of the variables  $x^j_{i,\alpha}: $ if $e_{ij}$ denotes
the $(i,j)-$matrix unit (i.e. the $N\times N$ matrix with 0 everywhere except the $i-$th row and $j-$th
column)  then $x^j_{i,\alpha} = {\tr}(e_{ij}M_{\alpha}).$ 
\end{example}

\subsubsection{Trace Poisson brackets from double brackets}
M. Van den Bergh defines a bracket operation on $A_N$ starting from a double Poisson structure on it.
We shall use Sweedler's notations: an element $\alpha\in A\otimes A$ shall be denoted by $\alpha = \alpha^{(1)}\otimes \alpha^{(2)}$
meaning that there is in fact a finite family $(\alpha^{(1)}_i,\alpha^{(2)}_i )$ in $A\times A$ such that $\alpha =\sum_i \alpha^{(1)}_i \otimes \alpha^{(2)}_i .$

\begin{thm}\cite{VdB1}
\begin{itemize}
\item  Given a double Poisson bracket on $A$ one defines a bracket $[-,-]:A\times A \to A$ 
\be\label{Lod}
[a,b] := \ldb  a,b\rdb^{(1)} \ldb a,b\rdb^{(2)} = \sum_i \ldb  a,b\rdb^{(1)}_i\, \ldb a,b\rdb^{(2)}_i
\ee
such that
\begin{itemize}
\item \eqref{Lod} satisfies the following derivation property:
$$[a,[b,c]] = [a,b],c]]+ [b,[a,c]];$$
\item The restriction $[-,-]:A_{\natural}\times A_{\natural} \to A_{\natural}$ defines on $A_{\natural}$ a Lie algebra structure:
\be\label{Lie}
[p(a),p(b)] := p([a,b]);
\ee
\item The map $[p(a), -] \in {\rm End}(A_{\natural})$ is induced by a derivation of  $A.$
\end{itemize}
\item Given a double Poisson bracket on $A$ one defines a Poisson structure on the representation variety $\Rep_N(A)$ i.e.
a Poisson bracket $\{-,-\}:A_N\times A_N \to A_N$ such that on generators  of $A_N$ (defined by elements $a$ and $b$ of $A$) we have
\be\label{PB}
\{a_{ij},b_{kl}\} := \ldb  a,b\rdb^{(1)}_{kj} \ldb a,b\rdb^{(2)}_{il}
\ee
\item The map $\tr : A_{\natural} \to A_N$ is a morphism of Lie algebras $A_{\natural}$ and $A_N = \C[\Rep_N (A)].$ Namely:
\be\label{traceP}
\{\tr p(a),\tr p(b)\} = \tr([p(a), p(b)]).
\ee
\end{itemize}
\end{thm}

\begin{defi}\label{doubbr} We shall refer to the Poisson brackets \eqref{PB} (following the suggestion in \cite{ORS2}) 
as {\it Trace Poisson brackets}. 
\end{defi}
It is an easy exercise to check that the Trace Poisson
 brackets are defined in fact  on the invariant part of $A_N$ or,
 more precisely, on conjugation classes
$\C[\Rep_N(A)]^{GL_N(\C)},$
where $GL_N(\C)$ acts on $A_N$ by conjugations. This is the unique Poisson structure on $\C[\Rep_N(A)]^{GL_N(\C)}$ such 
that \eqref{traceP} holds.

We shall be mostly interested by the Trace Poisson brackets induced by double brackets on a free associative algebra
such as considered in example \ref{free}.

\section{Associative Yang--Baxter equation}
We are now interested in Double bracket structures induced by endomorphisms $r\in {\rm End}(V\otimes V)$ (and not simply maps)
later identified with ``classical r-matrices''.
Associativity conditions on the Double Poisson algebra induces conditions of Yang--Baxter type on their structure constants
encapsulated in $r$. Note that the general situation of $r$ maps would gives rise to structures analog to the ``set-theoretical YB
equations'' studied in e.g. \cite{Caud}.

\subsection{AYBE and double Lie brackets}

T. Schedler \cite{Sch1} proposed the following existence criterion for double Lie brackets \eqref{DLB}: 
\begin{prop}\label{Dliecrit} Let $r\in {\rm End}(V\otimes V)$ defines the operation
\begin{equation}\label{dlie}
\ldb u,v \rdb := r(u\otimes v).
\end{equation}
This operation induces a double Lie bracket on $V$ iff  $r$ is {\rm skew-symmetric} and 
satisfies the {\bf Associative Yang--Baxter Equation (AYBE)} in $V\otimes V\otimes V:$
\begin{equation}\label{AYBE}
 AYBE(r):= \quad   r^{12}r^{13} - r^{23}r^{12} + r^{13}r^{23}=0,
\end{equation}
where, as usual, $r^{ij}$ acts in $V^{\otimes 3}$, non trivially on $(i,j)$ spaces and as identity elsewhere.
\end{prop}
Here the skew-symmetry of $r$ means that it satisfies  the condition
\begin{equation}\label{skewr}
r(v\otimes u) = -r(u\otimes v)^{o}, 
\end{equation}
which implies \eqref{skew}.

Conjugating \eqref{AYBE} by the permutation operator $P_{13}$ and using the skew-symmetry property of $r$
implies:
\begin{equation}
\label{AYBE2}
 AYBE^*(r) =\quad  r^{23}r^{12}+r^{31}r^{23}+r^{12}r^{31}=0\,.
\end{equation}
 
Now  $r\in {\rm End}(V)$
satisfies both \eqref{AYBE} and \eqref{AYBE2}. Such $r$ then satisfies the full {\bf Skew-Symmetric Classical Yang--Baxter Equation}:
\be
&&[r^{12}, r^{13}] + [r^{12}, r^{23}] + [r^{13}, r^{23}] = AYBE(r) - AYBE^*(r)  = \nonu
&&\qquad=  r^{12}r^{13} - r^{23}r^{12} + r^{13}r^{23}- (r^{23}r^{12}+r^{31}r^{23}+r^{12}r^{31} )= 0-0=0.
\ee 

The full classical Yang--Baxter equation for a non-skew symmetric $r$ matrix exhibits a 
different display of indices $32-13$ in the third term.

\subsubsection{AYBE and double Poisson brackets}
 
The following result of T. Schedler \cite{Sch1} is a direct corollary from the definitions and \eqref{Dliecrit} 
\begin{thm}
Let $A$ be any $\C-$algebra. An element $r \in {\rm End}_{\C}(A\otimes A)$ induces a 
double Poisson bracket iff $r$ is a skew element satisfying the AYBE 
and $r \in {\rm Der}_{A^{e}\otimes A^{e}}((A \otimes A)_{l,r}, (A \otimes A)_{in,out}).$
\end{thm}
\begin{remark}
We consider $(A \otimes A)_{l,r}$ as an $A^{e}\otimes A^{e}$-module by having the
first $A^{e}$ act on the first component, and the second on the second component:
$$
\big((u\otimes u^{o})\otimes (v\otimes v^{o})\big)(a\otimes b) = uau^{o}\otimes vbv^{o},
$$and consider  $(A \otimes A)_{in, out}$ as an $A^{e}\otimes A^{e}$-module by having the 
first $A^e$ act by inner multiplication and the second $A^e$ act by outer multiplication:
$$
\big((x\otimes x^{o})\otimes (y\otimes y^{o}) \big)(a\otimes b) = yax^{o}\otimes xby^{o}.
$$
\end{remark}

\section{The Trace Poisson brackets of the free associative algebras}

We shall from now on consider the situation of example \ref{free}. In addition we restrict ourselves to particular choices
of double brackets and their derived trace Poisson brackets, namely constant, linear and quadratic brackets.

\subsection{Three particular brackets}
\subsubsection{Constant, linear and quadratic brackets}
Let $A=\C<x_1,\ldots,x_m>$ be the free associative algebra. If the double brackets $\ldb x_\alpha, x_\beta \rdb$ between all 
generators are fixed, then the bracket between two arbitrary elements of $A$ 
is uniquely defined by identities (\ref{skew}) and (\ref{db}). 

The  constant, linear, 
and quadratic double brackets are defined respectively by 
\be \label{dconst}
&&\ldb x_\alpha,x_\beta\rdb = c_{\alpha\beta} 1\otimes 1, \mb{with} c_{\alpha,\beta}=-c_{\beta,\alpha},
 \\
&& \label{dlin}
\ldb x_\alpha,x_\beta\rdb = b_{\alpha \beta}^\gamma x_\gamma\otimes 1 - b_{\beta\alpha}^\gamma1\otimes x_\gamma,
\ee
and
\be
&&\label{dquad}
\ldb x_{\alpha}, x_{\beta}\rdb =r_{\alpha \beta}^{u v} \, 
x_u \otimes x_v+a_{\alpha \beta}^{v u} \, x_u x_v\otimes 1-a_{\beta \alpha}^{u v} \,1\otimes  x_v x_u,  
\ee
where
\begin{equation}\label{r1}
r^{\sigma \epsilon}_{\alpha\beta}=-r^{\epsilon\sigma}_{\beta\alpha}.
\end{equation}
 The summation with respect to repeated indexes is assumed.  

It is easy to verify that the bracket (\ref{dconst}) satisfies (\ref{db}) for any skew-symmetric tensor $c_{\alpha\beta}$.

The following observations of \cite{PVdW} gives us that
 the condition (\ref{db}) is equivalent for the bracket (\ref{dlin}) to the identity 
\be\label{r0}
b^{\mu}_{\alpha \beta} b^{\sigma}_{\mu \gamma}=b^{\sigma}_{\alpha \mu} b^{\mu}_{\beta \gamma},
\ee
which means that  $b^{\sigma}_{\alpha \beta}$
are structure constants of an associative algebra structure  on a vector space $V=\oplus_{i=1}^N\C x_i$.  

The corresponding statement for the quadratic bracket is more subtle. It was shown in \cite{ORS1} that
 the bracket (\ref{dquad}) satisfies (\ref{db}) iff the following relations hold:
\be\label{r2}
\begin{array}{ll}
r^{\lambda\sigma}_{\alpha\beta}
r^{\mu\nu}_{\sigma\tau}+r^{\mu\sigma}_{\beta\tau} r^{\nu\lambda}_{\sigma\alpha}+r^{\nu\sigma}_{\tau\alpha} r^{\lambda\mu}_{\sigma\beta}=0\,,\qquad
&a^{\sigma\lambda}_{\alpha\beta} a^{\mu\nu}_{\tau\sigma}=a^{\mu\sigma}_{\tau\alpha} a^{\nu\lambda}_{\sigma\beta}
\,,
\\[2.1ex]
a^{\sigma\lambda}_{\alpha\beta} a^{\mu\nu}_{\sigma\tau}=a^{\mu\sigma}_{\alpha\beta} r^{\lambda\nu}_{\tau\sigma}+a^{\mu\nu}_{\alpha\sigma}
r^{\sigma\lambda}_{\beta\tau}\,,
&a^{\lambda\sigma}_{\alpha\beta} a^{\mu\nu}_{\tau\sigma}=a^{\sigma\nu}_{\alpha\beta} r^{\lambda\mu}_{\sigma\tau}+a^{\mu\nu}_{\sigma\beta} r^{\sigma\lambda}_{\tau\alpha}\,.
\end{array}
\ee

\subsubsection{Trace brackets}
Let us specify the form of the trace Poisson brackets. 

We start with the constant trace algebra. It is somehow trivial, but fixes the notation:
\be
\label{constBr}
\lbrace x_{i,\alpha}^j, x_{i',\beta}^{j'}\rbrace = c_{\alpha\beta}\,\delta_{i}^{j'} \delta_{i'}^j 
\qquad\Leftarrow\qquad
\ldb x_{\alpha}, x_{\beta}\rdb = c_{\alpha\beta} 1\otimes 1 \,.
\ee
Here $i,j,i',j'$ are ``matrix'' indices running from $1$ to $N$ whereas $\alpha,\beta, \gamma$ are ``vector'' indices
running from $1$ to $m$. One naturally understands the set of variables $ x_{i,\alpha}^j$ as a $m$ vector-labeled set of $N \times N$
matrices. One immediately sees that the vector index is an extra feature of this trace algebra when compared to the usual
setting of classically integrable systems where the variables are encapsulated into a \textit{single} matrix. The vector
index shall later be denoted as ``flavor'' index using a transparent analogy with particle physics.

The linear trace algebra takes the form:
\be
\label{LinBr}
\lbrace x_{i,\alpha}^j, x_{i',\beta}^{j'}\rbrace &=& b_{\alpha\beta}^{\gamma}x_{i,\gamma}^{j'} \delta_{i'}^j - 
b_{\beta\alpha}^{\gamma}x_{i',\gamma}^j \delta_i^{j'}
\quad\Leftarrow\quad
\ldb x_{\alpha}, x_{\beta}\rdb = b_{\alpha\beta}^{\gamma} x_{\gamma}\otimes 1 - b_{\beta\alpha}^{\gamma} 1\otimes x_{\gamma}\,.\qquad\quad
\ee

The trace Poisson bracket corresponding to the general quadratic double Poisson bracket (\ref{dquad})  
can be defined on  $\C[{\rm Rep}_N (A)]$ in the following way \cite{ORS2}: 
\be\label{Poisson}
\{x^j_{i,\alpha},x^{j^{\prime}}_{i^{\prime},\beta}\}=
r^{\gamma\epsilon}_{\alpha\beta}x^{j^{\prime}}_{i,\gamma}x^j_{i^{\prime},\epsilon}+
a^{\gamma\epsilon}_{\alpha\beta}x^k_{i,\gamma}x^{j^{\prime}}_{k,\epsilon}\delta^j_{i^{\prime}}-
a^{\gamma\epsilon}_{\beta\alpha}x^k_{i^{\prime},\gamma}x^{j}_{k,\epsilon}\delta^{j^{\prime}}_i\,,
\ee
where $x^j_{i,\alpha}$ are entries of the matrix $x_{\alpha}$ and $\delta^{j}_i$ is the 
Kronecker delta-symbol. Relations  (\ref{r1}) and (\ref{r2}) hold iff (\ref{Poisson}) 
is a Poisson bracket.

It is finally interesting to note that these brackets also take a Hamiltonian form following:
\begin{remark}\label{hamilt}
Using the observation at the end of the example \ref{free} one writes \eqref{Poisson} as
\be\label{PoissonTr}
\{x^j_{i,\alpha},x^{j^{\prime}}_{i^{\prime},\beta}\}=\{{\tr}(e_{ij}M_{\alpha}), {\tr}(e_{i^{\prime}j^{\prime}}M_{\beta})\}.
\ee

The constant bracket can be rewritten as 
\be
\{ x_{\alpha}, x_{\beta}\} = {\tr}(e_{ij} \,c_{\alpha,\beta}\,e_{i^{\prime}j^{\prime}})\,.
\ee

The linear Trace-Poisson brackets then  read:
\be\label{PoissonTr2}
\{x^j_{i,\alpha},x^{j^{\prime}}_{i^{\prime},\beta}\}
={\tr}(e_{ij} \Theta_{\alpha,\beta}(e_{i^{\prime}j^{\prime}}))\,.
\ee
Following \cite{MikSok}, we have introduced the Hamiltonian operator $\Theta\in {\mathcal A}(A) \otimes {\mathbb M}_N(\C)$ for
the algebra ${\mathcal A}(A)$ generated by left- and right multiplications in $A=\C<x_1,...,x_m>:$
\be\label{LinHam}
\Theta_{\alpha,\beta} = b^{\sigma}_{\alpha \beta} L_{x_{\sigma}}- b^{\sigma}_{\beta \alpha} R_{x_{\sigma}},
\ee
where $b^{\sigma}_{\alpha \beta}$ are structure constants of an associative algebra as above in \eqref{dlin}, and
\be
L_{x_{\alpha}}:\ \left\{\begin{array}{ccl}
A&\to& A\\
y&\to& L_{x_{\alpha}}(y)=x_{\alpha}y\end{array}\right.
\quad \text{and}\quad 
R_{x_{\beta}} :\ \left\{\begin{array}{ccl}
A&\to& A\\
y&\to& R_{x_{\beta}}y =y{x_{\beta}}.\end{array}\right.
\ee

It is not difficult to write the Hamiltonian operator for the quadratic trace-Poisson brackets \eqref{dquad}.
\be
\Theta_{\alpha,\beta} = a^{\sigma \epsilon}_{\alpha \beta} L_{x_\sigma} L_{x_\epsilon} - a^{\epsilon\sigma}_{\beta \alpha} R_{x_\sigma}
R_{x_\epsilon} + r^{\sigma \epsilon}_{\alpha \beta} L_{x_\sigma} R_{x_\epsilon},
\ee
where $a^{\sigma \epsilon}_{\alpha \beta}$ and $ r^{\sigma \epsilon}_{\alpha \beta}$ satisfy the relations \eqref{r1}.
\end{remark}

Our main purpose now is to reformulate the linear and quadratic trace Poisson algebra 
in a fully algebraic notation involving the
relevant generalization of a classical linear or quadratic  $r$ matrix structure.

\subsection{Linear trace-brackets: the r-matrix formulation}

The linear trace-Poisson algebra is re-expressed using a matrix-Poisson formula, with notations derived from
the canonical classical $r$-matrix formalism \cite{STS} but augmented by the flavor indices. 
Accordingly one introduces two auxiliary vector spaces $\mathbb C^N$ and $\mathbb C^m$ and define the following objects
embedded in the general tensorized structure 
$\MM_m(\CC)\otimes \MM_N(\CC)\otimes \CC^m\otimes \MM_N(\CC)$, where $\MM_m(\CC)={\rm End}(\mathbb C^m)$:
\be
\label{notation1}
B_{12} &=&\sum_{\alpha,\beta,\gamma =1}^m \sum_{ij=1}^N b_{\alpha\beta}^{\gamma}
\,e_{\alpha\gamma}\otimes e_{ij}\otimes e_{\beta}\otimes e_{ji}\in 
\MM_m(\CC)\otimes \MM_N(\CC)\otimes \CC^m\otimes \MM_N(\CC),\quad\\
B_{21} &=& \sum_{\alpha,\beta,\gamma =1}^m \sum_{ij=1}^N b_{\alpha\beta}^{\gamma}\, e_{\beta}\otimes 
e_{ij}\otimes e_{\alpha\gamma}\otimes e_{ji}\in \CC^m\otimes \MM_N(\CC)\otimes 
\MM_m(\CC)\otimes \MM_N(\CC),\qquad\\
X&=&\sum_{\alpha =1}^m \sum_{ij=1}^N x_{i\alpha}^j e_{\alpha}\otimes e_{ji}\in {\mathbb C^m}\otimes \MM_N(\CC),\
\label{notation2}\\
X_1 &=& X\otimes {\mathbb I_m}\otimes{\mathbb I_N}\in  {\mathbb C^m}\otimes \MM_N(\CC)\otimes\MM_m(\CC)\otimes 
\MM_N(\CC),\\
X_2 &=& \sum_{\alpha =1}^m \sum_{ij=1}^N  x_{i\alpha}^j \,{\mathbb I_m}\otimes{\mathbb I_N}\otimes  
e_{\alpha}\otimes e_{ji}\in {\mathbb C^m}\otimes \MM_N(\CC)\otimes\MM_m(\CC)\otimes \MM_N(\CC).
\ee
$\II$ denotes the identity operator in the corresponding vector space. 

We define:
\be
\label{brack1}
\lbrace X_1 \stackrel{\otimes}{,} X_2 \rbrace \equiv \lbrace x_{i,\alpha}^j, x_{i',\beta}^{j'}\rbrace\,
 e_{\alpha}\otimes 
e_{ji}\otimes e_{\beta}\otimes e_{j'i'} \in {\mathbb C^m}\otimes \MM_N(\CC)\otimes \CC^m\otimes 
\MM_N(\CC)\quad
\ee
which yields:
\be
\label{brack2}
\lbrace X_1 \stackrel{\otimes}{,} X_2 \rbrace = b_{\alpha\beta}^{\gamma}\,x_{i,\gamma}^{j'}\,e_{\alpha}\otimes e_{ji}\otimes e_{\beta}\otimes e_{j'j} - 
b_{\beta\alpha}^{\gamma}\,x_{i',\gamma}^j \,e_{\alpha}\otimes e_{ji}\otimes e_{\beta}\otimes e_{ii'}.
\ee

One easily verifies:
\be
B_{12}X_1 &= &
b_{\alpha\beta}^{\gamma}x_{i,\gamma}^{j}\,e_{\alpha}\otimes e_{i'i}\otimes e_{\beta}\otimes e_{ji'}\,,
\\
B_{21}X_2 &=& 
b_{\alpha\beta}^{\gamma}x_{i,\gamma}^{j}\,e_{\beta}\otimes e_{ji'}\otimes e_{\alpha}\otimes e_{i'i}\,.
\ee
One then deduces the matrix form of the trace Poisson algebra :
\begin{prop}
\label{Rmat}
The notation $\lbrace X_1 \stackrel{\otimes}{,} X_2 \rbrace = B_{12}X_1 - B_{21}X_2$ reproduces the relations of the linear brackets for
$$X=\sum_{\alpha =1}^m \sum_{ij=1}^N x_{i\alpha}^j e_{\alpha}\otimes e_{ji},\quad B_{12} =\sum_{\alpha,\beta,\gamma =1}^m \sum_{ij=1}^N 
b_{\alpha\beta}^{\gamma}\,e_{\alpha\gamma}\otimes e_{ij}\otimes e_{\beta}\otimes e_{ji} \sim b_{12}\otimes P,$$
where $P\in \MM_N(\CC)\otimes \MM_N(\CC)$ is the flip operator and $b_{12} = b_{\alpha\beta}^{\gamma}\,e_{\alpha\gamma}\otimes e_{\beta}$.

 We have used the symbol $\sim$ for an equality valid up to re-ordering in the tensor product of spaces.
\end{prop}

The proof is by direct identification. 
We now write the associativity condition for the trace-Poisson algebra in terms of these new notations:

\begin{prop}
\label{ass}
The algebra structure on the $\mathbb C-$vector space $V=<e_1,..,e_m>$ given by the tensor $b_{\alpha\beta}^{\gamma}:$
$$e_{\alpha}e_{\beta} = b_{\alpha\beta}^{\gamma}e_{\gamma}$$ satisfies  the associativity constraint iff
$$b_{\alpha\beta}^{\mu}b_{\alpha\mu}^{\sigma} = b_{\alpha\mu}^{\sigma}b_{\beta\gamma}^{\mu}\iff b_{12}b_{13}=b_{23}b_{12}\iff B_{12}B_{13}=B_{23}B_{12}.$$
\end{prop}

The proof is immediate. Indeed, 
$$
b_{12}b_{13}= b_{\alpha\beta}^{\gamma}b_{\gamma\beta'}^{\gamma'}e_{\alpha\gamma'}\otimes e_{\beta}\otimes e_{\beta'}
=b_{23}b_{12}.
$$
For the second equivalence it is enough to observe that
$$B_{12}B_{13}\sim b_{12}b_{13}\otimes P_{12}P_{13},\quad B_{23}B_{12}\sim b_{23}b_{12}\otimes P_{23}P_{12}$$ but
 $P_{23}P_{12}=P_{12}P_{13}$
which proves the claim.

\subsection{Quadratic trace brackets: the r-matrix formulation}

We recall the form of the quadratic Poisson brackets \eqref{Poisson}:
\begin{eqnarray}
\label{QuadBr}
\lbrace x_{i,\alpha}^j, x_{i',\beta}^{j'}\rbrace& =& r_{\alpha\beta}^{\gamma\epsilon}x_{i,\gamma}^{j'} x_{i',\epsilon}^j + 
a_{\alpha\beta}^{\gamma\epsilon}x_{i,\gamma}^{k} x_{k,\epsilon}^{j'}\delta_{i'}^{j} - 
a_{\beta\alpha}^{\gamma\epsilon}x_{i',\gamma}^{k} x_{k,\epsilon}^{j}\delta_{i}^{j'}\,,\\
\ldb x_{\alpha}, x_{\beta}\rdb& =& r_{\alpha\beta}^{\gamma\epsilon} x_{\gamma}\otimes x_{\epsilon} + 
a_{\alpha\beta}^{\epsilon\gamma}x_{\gamma}x_{\epsilon}\otimes 1 -a_{\beta\alpha}^{\gamma\epsilon} 1\otimes x_{\epsilon}x_{\gamma}\,.
\end{eqnarray}

Using the same embedding as before one introduces the following objects:
\be
\label{notation3}
\fr_{12} &=&\sum_{\alpha,\beta,\gamma,\epsilon}^m\sum_{ij=1}^N r_{\alpha\beta}^{\gamma\epsilon}e_{\alpha\gamma}\otimes e_{ij}\otimes 
e_{\beta\epsilon}\otimes e_{ji}\
\sim\ r_{12}\otimes P,\\
\label{notation4}
\fa_{12} &=&\sum_{\alpha,\beta,\gamma,\epsilon}^m\sum_{ij=1}^N a_{\alpha\beta}^{\gamma\epsilon}e_{\alpha\gamma}\otimes e_{ij}\otimes 
e_{\beta\epsilon}\otimes e_{ji}
\sim a\otimes P. 
\ee
One then has
\be\label{notation3b}
\fr_{12}X_1X_2 
&=& r_{\alpha\beta}^{\gamma\epsilon}x_{i'\gamma}^{j} x_{k\epsilon}^i\,
 e_{\alpha}\otimes e_{ii'}\otimes e_{\beta}\otimes e_{jk},
\\
X_2^t \fa_{12}X_1 
&=& a_{\alpha\beta}^{\gamma\epsilon}x_{i'\gamma'}^{j} x_{j\epsilon}^l
\, e_{\alpha}\otimes e_{ii'}\otimes e_{\beta}^t\otimes e_{li},
\\
\label{notation5}
X_1^t \fa_{21}X_2 
&=& a_{\alpha\beta}^{\gamma\epsilon}x_{i\epsilon}^{j'} x_{k\gamma}^i\, e_{\beta}^t\otimes e_{j'j}\otimes e_{\alpha}\otimes e_{jk}.
\ee

It is crucial to emphasize here that the transposition $x^t$ is a \textit{partial} transposition taking place in
the flavor space $\mathbb C^m.$ A ``transposed'' vector is of course now a co-vector or a linear form, identified
by canonical duality.

The Poisson brackets can be expressed as
\be
\lbrace X_1 \stackrel{\otimes}{,} X_2 \rbrace &=& r_{\alpha\beta}^{\gamma\epsilon}x_{i\gamma}^{j'} x_{i'\epsilon}^j e_{\alpha}\otimes e_{ji}\otimes 
e_{\beta}\otimes e_{j'j} \nonu
&&+ a_{\alpha\beta}^{\gamma\epsilon}x_{i\gamma}^{k} x_{k\epsilon}^{j'}e_{\alpha}\otimes e_{ji}\otimes e_{\beta}\otimes e_{j'j}  - 
a_{\beta\alpha}^{\gamma\epsilon}x_{i'\gamma}^{k} x_{k\epsilon}^{j}e_{\alpha}\otimes e_{ji}\otimes e_{\beta}\otimes e_{ii'}.
\qquad \label{brack2b}
\ee

The new $r$-matrices $R$ and $A$ only carry non-trivial indices of flavor type. 
In other words the non-trivial contributions to Poisson structure arise solely between any two $X$ matrices with different 
flavors whilst the same-flavor Poisson structure is trivial. This is again an important distinction with respect to 
standard quadratic Poisson structure formulation.

The formulas \eqref{brack2} and \eqref{notation3}-\eqref{notation5} finally result in the following
\begin{prop}\label{quadrback}
The quadratic Poisson brackets \eqref{QuadBr} can be rewritten as 
\be
\lbrace X_1 \stackrel{\otimes}{,} X_2 \rbrace = \fr_{12}X_1X_2 + (X_2^t \fa_{12}X_1)^{t_2} - (X_1^t \fa_{21}X_2)^{t_1}
\ee 
where
\be
X &=& \sum_{\alpha =1}^m \sum_{ij=1}^N x_{i\alpha}^j \,e_{\alpha}\otimes e_{ji}\,,
\\
\fr_{12} &=&\sum_{\alpha,\beta,\gamma,\epsilon}^m\sum_{ij=1}^N r_{\alpha\beta}^{\gamma\epsilon}
\,e_{\alpha\gamma}\otimes e_{ij}\otimes e_{\beta\epsilon}\otimes e_{ji}\
,\\
\fa_{12} &=&\sum_{\alpha,\beta,\gamma,\epsilon}^m\sum_{ij=1}^N a_{\alpha\beta}^{\gamma\epsilon}e_{\alpha\gamma}\otimes e_{ij}\otimes e_{\beta\epsilon}\otimes e_{ji}\,.
\ee
\end{prop}
The properties of $r$ and $a$ implies the following relations for $\fr$ and $\fa$:
$$r_{12}=-r_{21}\Rightarrow \fr_{12}=-\fr_{21},\quad a_{12}= a_{21}\Rightarrow \fa_{12}= \fa_{21}.$$

\subsection{Classical YBE for the full r-matrix structure ($\fr, \fa$) }
Consider now the Yang--Baxter equations for the full structure matrices $\fr$ and $\fa$ deduced from the compatibility 
equations for the components $r,a$ previously obtained. We first have:

\begin{prop}\label{cl-ass}
If $r$ is a skew-symmetric solution of the AYBE \eqref{r2} then $\fr$, as defined in proposition \ref{quadrback}, 
 is a solution of the classical Yang--Baxter equation 
\be
\label{CYBE-1}
[\fr_{12}, \fr_{13}+ \fr_{23}] +[\fr_{13},\fr_{23}] = 0\,.
\ee

\end{prop}

It is indeed easy to see that :
\beano
[\fr_{12}, \fr_{13}+ \fr_{23}] +[\fr_{13},\fr_{23}] &=& (r_{12}r_{13} - r_{23}r_{12}+r_{13}r_{23})\otimes P_{12}P_{13}
\\
&& +
( r_{12}r_{23}- r_{13}r_{12}- r_{23}r_{13})\otimes P_{12}P_{23} =0\,.
\eeano

When a classical $r$-matrix is not skew-symmetric it obeys a generalized version of this better-known CYB equation \cite{STS,Mail}.
It is however more relevant, in particular with respect to quantization issues, to formulate the CYB conditions for a 
split pair, here $(\fr,\fa)$ where one assumes in addition that
$\fr$ is skew-symmetric and $\fa$ is symmetric (i.e. $\fa_{12} = \fa_{21 }$). The skew-symmetric part $\fr$ as we have just establish,
obeys the canonical skew-symmetric
YB equation \eqref{CYBE-1}. The other condition now is the adjoint $(\fr,\fa)$ equation \cite{Mail} which reads: 
\be
[\fr_{12}, \fa_{13}+ \fa_{23}] +[\fa_{13}, \fa_{23}] &=& r_{12}a_{13}\otimes P_{12}P_{13} + r_{12}a_{23}\otimes 
P_{12}P_{23}+a_{13}a_{23}\otimes P_{13}P_{23}\nonu
&&- a_{13}r_{12}\otimes P_{13}P_{12} - a_{23}r_{12}\otimes P_{23}P_{12} - a_{23}a_{13}\otimes P_{23}P_{13}
\nonu
&=&0.\label{CYBE-2}
\ee
Note that the sum $\fr+\fa$ then obeys the non-skew-symmetric
classical YB equation given some general properties of $\fr, \fa$ which we shall not detail here.

One now establishes easily that:
\begin{prop}\label{cl-ass-2}
If $(r,a)$ is a solution of the AYBE relations \eqref{r1} such that $a_{12}=a_{21}$ then $(\fr,\fa)$, 
as defined in proposition \ref{quadrback},  
is a solution of the adjoint classical Yang--Baxter equation \eqref{CYBE-2}.
\end{prop}

This is a direct consequence of the formula
\beano
[\fr_{12}, \fa_{13}+ \fa_{23}] +[\fa_{13},\fa_{23}] &=& (r_{12}a_{13} - a_{23}r_{12}+a_{13}a_{23})\otimes P_{12}P_{13}
\\
&& +
( r_{12}a_{23}- a_{13}r_{12}- a_{23}a_{13})\otimes P_{12}P_{23}
\eeano
and the implication:
$$
r_{13}a_{12}-a_{32}r_{13} = a_{32}a_{12}\Rightarrow r_{12}a_{13}-a_{23}r_{12} = a_{23}a_{13}
$$
(via the permutation of spaces $2\to 3$).
We use also $a_{12}a_{31}=a_{31}a_{12}$ and $a_{12}=a_{21}$ which imply $a_{12}a_{31}=a_{13}a_{21}.$
Hence
$a_{23}a_{13}=a_{32}a_{13}=a_{31}a_{23}.$

\subsection{Reflection Algebra Poisson brackets}
The formulation of quadratic trace-Poisson structure still lacks one type of term as can be seen from
the famous general quadratic Poisson bracket ansatz of L. Freidel and J.-M. Maillet \cite{FM}
\be
\label{mail}
\lbrace l_1 \stackrel{\otimes}{,} l_2\rbrace = {\tilde a_{12}}l_1 l_2 + l_1{\tilde b_{12}}l_2 - l_1 l_2{\tilde d_{12}} + l_2{\tilde c_{12}}l_ 1\,.
\ee
To get a full quadratic classical Poisson structure, hereafter denoted Reflection Algebra (R.A.), it then appears that we need to add to our formulas some analogue  
of the last term $\tilde d$. This is achieved by introducing the transposed matrix of $\fr$ defined by:
$$
{\tilde \fr_{12}}= {\fr_{12}^{t_{12}}} = \sum_{\alpha,\beta,\gamma,\epsilon}^m\sum_{ij=1}^N 
r_{\alpha\beta}^{\gamma\epsilon}e_{\gamma\alpha}\otimes e_{ij}\otimes e_{\epsilon\beta}\otimes e_{ji}\,.
$$
Then one  has the identification:
$$(X_2^t X_1^t {\tilde \fr_{12}})^{t_{12}} = \fr_{12}X_1X_2$$ 
so that we end up with the fully quadratic formulation a la Freidel--Maillet.

\begin{prop}\label{REAbrack}
The reflection classical trace Poisson algebra is defined as:
\be
\label{QCTA}
\lbrace X_1 \stackrel{\otimes}{,} X_2 \rbrace = \frac{1}{2}\fr_{12}X_1X_2 -  \frac{1}{2}(X_2^t X_1^t 
{\tilde \fr_{21}})^{t_{12}} + (X_2^t \fa_{12}X_1)^{t_2} - (X_1^t \fa_{21}X_2)^{t_1}.
\ee
We have used the notation of proposition \ref{quadrback}.
 \end{prop}
This now clearly suggests a consistent quantization of the trace Poisson algebra as a reflection algebra type structure.

\section{Quantum Reflection trace algebra}
\subsection{Quantisation of the trace-Poisson algebra}
We now move to the formulation of a quantization for these Poisson algebras. We first consider with the quadratic trace-Poisson 
algebra discussed above and propose a quantized form. We then introduce a general quantum quadratic algebra and define its associated Yang--Baxter equations,
using a generalization of the bivector formulation of Freidel--Maillet.

 The  quantization of a  single-flavor general quadratic Poisson bracket has been done in \cite{FM}. Here, we need to take
into account the delicate issue of transposition with respect to the flavor indices, as it occurred in the previously derived
formulae. By analogy, we come up with the following proposition of a quantum quadratic trace algebra:
\begin{prop}
\label{prop:lim-class}
Let $R$ and $A$ be two matrices acting on the tensor product of two copies of an auxiliary space $V=\CC^N\otimes\CC^m$. 
The space $\CC^N$ (resp. $\CC^m$) will be called the color (resp. flavor) space.

We define an associative algebra $\cA$ through the following relation
\be
\label{QQTA}
({R_{12}}({K_1}^t{A_{21}}{K_2})^{t_1}) ^{t_2}= (({K_2}^{t}{A_{12}}{K}_1)^{t_1}{R_{12}}^{t_1t_2})^{t_1},
\ee
where the transposition acts solely on the flavor indices, and $K\in\cA\otimes End(\CC^N)\otimes\CC^m$.

Then $\cA$ is a quantization of the trace Poisson algebra defined in proposition \ref{REAbrack}.
\end{prop}
Note that the following relation, deduced from \eqref{QQTA} by exchanging the auxiliary spaces 1 and 2,
\be\label{QQTA2}
(R_{21}(K_2^t A_{21})^{t_2}K_1)^{t_1} = (K_1^t(A_{21}K_2)^{t_2}R_{21}^{t_1t_2})^{t_2}
\ee
is a priori \underline{not}\footnote{Of course, this is true 
only when the flavor space is not trivial: when the flavor space is one-dimensional, 
the two relations become equivalent.} equivalent to \eqref{QQTA} and must also be considered simultaneously.

Consistency of proposition \ref{prop:lim-class} is now proved by defining a quasi-classical limit, assuming the existence
of the following $\hbar$ expansions:

$${R} ={\mathbb I} - \hbar \fr + \dots, \quad {A} ={\mathbb I} + \hbar \fa + \dots, \quad { K} =X + \hbar (Y);$$
$$[X_1,X_2] = \hbar\lbrace  X_1,X_2\rbrace + \ldots$$

There is no contribution to order $\hbar^0$. The term in $\hbar^1$ reads:
 $$
 \lbrace X_1\stackrel{\otimes}{,}X_2^t\rbrace = (\fr_{12}X_1X_2)^{t_2} +(X_1^t\fa_{21}X_2) ^{t_{12}}-(X_2^t\fa_{12}X_1) - (X_2^tX_1^t{\tilde \fr_{12}})^{t_1}\,.
 $$
If we apply now the transposition $()^{t_2}$ we get:
$$
 \lbrace X_1\stackrel{\otimes}{,}X_2\rbrace = (\fr_{12}X_1X_2) +(X_1^t\fa_{21}X_2) ^{t_1}-(X_2^t\fa_{12}X_1)^{t_2} - (X_2^tX_1^t{\tilde \fr_{12}})^{t_{12}}
$$
which is exactly the classical quadratic trace algebra. We thereby prove the consistency of the choice of \eqref{QQTA}
as a quantization of \eqref{QCTA}. 

\medskip

Proposition \ref{prop:lim-class} suggests to introduce what appears as the most general quadratic quantum exchange algebra
of trace-type, introducing four a priori distinct structure constant matrices $A,B,C,D$ in a notation
directly borrowed from \cite{FM}.

\begin{defi}
The quantum trace Reflection Algebra relations read:
\be\label{etoile1}
(A_{12}(K_1^tB_{12})^{t_1}K_2)^{t_2} = (K_2^t(C_{12}K_1)^{t_1}D_{12})^{t_1}\,.
\ee
Again, we recall that each of the auxiliary spaces $1$ and $2$ is itself  a tensor space of a flavor and a color space. 
The transposition acts solely on the flavor indices. As before, the 
following relation, deduced from \eqref{etoile1} by exchanging the auxiliary spaces 1 and 2,
\be\label{etoile2}
(A_{21}(K_2^tB_{21})^{t_2}K_1)^{t_1} = (K_1^t(C_{21}K_2)^{t_2}D_{21})^{t_2}
\ee
is a priori \underline{not}\footnote{Again, this is true 
only when the flavor space is not trivial.} equivalent to \eqref{etoile1} and must be considered simultaneously.
\end{defi}

\subsection{Freidel--Maillet R.A. formulation and YB equation}
In order to derive sufficient conditions for associativity of this R.A. we need to introduce an alternative representation.
For single-flavor R.A., it was 
originally proposed in \cite{FM} and possibly related (in this context) to the interpretation of 
R.A. as twists of a tensor product of several quantum algebras \cite{DKM}. This representation interprets the $K$ matrices as partially
transposed bivectors. It yielded
a completely bivector form for the R.A. of Z.F. algebra type: $RKK = KK$ \cite{ZF}. 

In our context, we must bi-vectorialize the color space $End(\CC^N)$ of $K$, which yields the following proposition:
\begin{prop}
The quantum reflection trace algebra can be reformulated as:
\be\label{FMform}
{\cR}_{11',22'}^{\sun,\sdx}\,\cK^{\sun}_{11'}\cK^{\sdx}_{22'}= \cK^{\sdx}_{22'}\cK^{\sun}_{11'}
\mb{where} {\cR}_{11',22'}^{\sun,\sdx}=(C_{12'}^{T_{2'}})^{-1}(D_{1'2'}^{T_{1'}T_{2'}})^{-1}A_{12}B_{1'2}^{T_{1'}}\,,
\ee
As a convention, $11'$ and $22'$ denote the color spaces that are bivectorialized, 
while $\sun$  and $\sdx$ label the flavor spaces.
 The transpositions are defined as $T_{1'}\equiv t_{1'}t_{\sun}$ and $T_{2'}\equiv t_{2'}t_{\sdx}$. 
 
 For the sake of simplicity we have omitted the flavor labels $\sun$ and $\sdx$ in the matrices $A$, $B$, $C$, $D$.

\end{prop}

This is easily seen by expanding the relation on the canonical basis. The matrices $A$, $B$, $C$, $D$ are generically expanded as
$$
M = \sum_{i,j,k,l=1}^N \sum_{\alpha,\beta,\gamma,\delta=1}^m M_{\alpha\beta,\gamma\delta}^{ij,kl}\ 
e_{ij}\otimes e_{\alpha\beta}\otimes e_{kl}\otimes e_{\gamma\delta}
$$
and $\cK$ is obtained from $K$ as 
$$
K = \sum_{i,j=1}^N \sum_{\alpha=1}^m K_{\alpha}^{ij}\ e_{ij}\otimes e_{\alpha}
\quad\Rightarrow\quad\cK = \sum_{i,j=1}^N \sum_{\alpha=1}^m \cK_{\alpha}^{i,j}\ e_{i}\otimes e_{j}\otimes e_{\alpha}
\mb{with}K_{\alpha}^{ij}=\cK_{\alpha}^{i,j} .
$$
R.A. projected on $e_{i}\otimes e_{q}\otimes e_{\mu}\otimes e_{k}\otimes e_{s}\otimes e_{\nu}$ now reads:
\be
\label{REAind}
\sum_{j,l,n,p=1}^N \sum_{\alpha,\beta,\gamma,\delta=1}^m
A_{\mu\beta,\nu\gamma}^{ij,kl}B_{\alpha\beta,\gamma\delta}^{nq,lp}\cK_{\alpha}^{j,n}\cK_{\delta}^{p,s} = 
\sum_{j,n,u,r=1}^N \sum_{\alpha,\beta,\gamma,\delta=1}^m
D_{\mu\alpha,\nu\delta}^{qn,us}C_{\alpha\beta,\gamma\delta}^{ij,ru}\cK_{\gamma}^{k,r}\cK_{\beta}^{j,n}.
\ee
A careful reinterpretation of the indices in \eqref{REAind} allows us to give it a matricial form:
\be
A_{12}B_{1'2}^{T_{1'}}\cK^{\sun}_{11'}\cK^{\sdx}_{22'}= D_{1'2'}^{T_{1'}T_{2'}}\,C_{12'}^{T_{2'}}\,
\cK^{\sdx}_{22'}\cK^{\sun}_{11'}
\ee
Multiplying by the inverse matrices of $D^{T_{1'}T_{2'}}_{1'2'}$ and $C^{T_{2'}}_{12'}$,
 one gets the  Z.F. algebra type relation \eqref{FMform}.

Note that $\cR$ is now expanded on the canonical basis as
\be
{\cR}_{11',22'}^{\sun,\sdx}=\sum_{\atopn{i,j',p,q'}{r,s',k,l'}=1}^N\ \sum_{\alpha,\alpha',\gamma,\gamma'=1}^m
\cR_{\alpha\alpha',\gamma\gamma'}^{ij',pq';rs',kl'}\,
\underbrace{e_{ij'}}_1\otimes \underbrace{e_{pq'}}_{1'}\otimes \underbrace{e_{\alpha\alpha'}}_{\sun}\otimes \underbrace{e_{rs'}}_{2}\otimes \underbrace{e_{kl'}}_{2'}\otimes \underbrace{e_{\gamma\gamma'}}_{\sdx}
\ee
with
\be
\cR_{\alpha\alpha',\gamma\gamma'}^{ij',pq';rs',kl'}=
\sum_{j,l,q,s=1}^N\sum_{\atopn{\beta,\beta',\beta"}{\delta,\delta',\delta"}=1}^m
\wt C_{\alpha\beta,\gamma\delta}^{ij,kl}\,\wt D_{\beta\beta',\delta\delta'}^{pq,ll'}
\,A_{\beta'\beta",\delta'\delta"}^{jj',rs}\,B_{\alpha'\beta",\delta"\gamma'}^{qq',ss'}
\ee
where $\wt C_{\alpha\beta,\gamma\delta}^{ij,kl}$ corresponds to the expansion of $(C_{12'}^{T_{2'}})^{-1}$
\be
\sum_{j,l=1}^N\sum_{\beta,\delta=1}^m
\wt C_{\alpha\beta,\gamma\delta}^{ij,kl}\, C_{\beta\beta',\delta'\delta}^{jj',l'l}\ =\ \delta_{\alpha\beta'}\,
\delta_{\gamma\delta'}\,\delta^{ij'}\,\delta^{kl'}\,,
\ee
and $\wt D_{\alpha\beta,\gamma\delta}^{ij,kl}$ to  the expansion of $(D_{12'}^{T_{1'}T_{2'}})^{-1}$
\be
\sum_{j,l=1}^N\sum_{\beta,\delta=1}^m
\wt D_{\alpha\beta,\gamma\delta}^{ij,kl}\, D_{\beta'\beta,\delta'\delta}^{j'j,l'l}\ =\ \delta_{\alpha\beta'}\,
\delta_{\gamma\delta'}\,\delta^{ij'}\,\delta^{kl'}\,.
\ee

\subsection{Consistency conditions}
\subsubsection{Conditions of unitarity}

So-called ``Unitarity conditions'' follow from requiring that the R.A. and its rewriting by exchange of the auxiliary space
labels $1$ and $2$ have the same content. Indeed in the $RTT = TTR$ case this condition only yields
unitarity conditions on the quantum $R$ matrix hence the name. 

The Freidel--Maillet formulation now provides a simple way to get this unitary condition. 
Comparing both writings leads immediately to
\be
{\cR}^{\sun,\sdx}_{11',22'}\,{\cR}^{\sdx,\sun}_{22',11'}\,={\mathbb I},
\ee
where ${\cR}^{\sun,\sdx}_{11',22'}$ is given in \eqref{FMform}. 
Then,  one  gets the following sufficient conditions:
\be\label{unitar}
C^{\sun,\sdx}_{12} = B^{\sdx,\sun}_{21}\mb{and} 
\big((D^{\sun,\sdx}_{1'2'})^{T_{1'}T_{2'}}\big)^{-1} A^{\sun,\sdx}_{12}
=\,(A^{\sdx,\sun}_{21})^{-1}\,(D^{\sdx,\sun}_{2'1'})^{T_{1'}T_{2'}}.
\ee

\subsubsection{Conditions of associativity (YBE)}
Let us now examine the associativity property of the R.A.
\begin{thm}
A sufficient condition
for associativity of the R.A. is  given by the Quantum Yang--Baxter equation for ${\cR}$:
\be\label{YB}
{\cR}^{\sun,\sdx}_{11',22'}\,{\cR}^{\sun,\str}_{11',33'}\,{\cR}^{\sdx,\str}_{22',33'}
={\cR}^{\sun,\sdx}_{22',33'}\,{\cR}^{\sun,\str}_{11',33'}\,{\cR}^{\sdx,\str}_{11',22'}\,,
\ee
where ${\cR}^{\sun,\sdx}_{11',22'}$  is the R-matrix introduced in \eqref{FMform}.
\end{thm}

Due to the mixing of flavor indices between $A,B,C,D$ in the definition of ${\cR}^{\sun,\sdx}_{11',22'}$ it is
more delicate to separate \eqref{YB} into four distinct QYB equations (schematically written as $AAA$, $ABB$, $DCC$ and $DDD$) as was done in the Freidel--Maillet R.A. We give below some examples where it can nevertheless be done.

\subsubsection{Sufficient conditions for consistency}
In this section, we shall assume that the flavor and color indices are completely decoupled:
$$
M^{\sun,\sdx}_{11',22'}=M_{\sun,\sdx}\otimes\,M_{11',22'} \mb{for} M=A,B,C,D\,.
$$   

Unitarity condition for $\cR^{\sun,\sdx}_{11',22'}$ is  fulfilled when:
\be
&&C_{12} = B_{21}\mb{;} 
D_{1'2'}\,D_{2'1'}=\II \mb{;} A_{12}\,A_{21}=\II,\\
&&C_{\sun,\sdx} = B_{\sdx,\sun}\mb{;} 
\big(D_{\sun,\sdx}^{t_{\sun}t_{\sdx}}\big)^{-1} A_{\sun,\sdx}
=\,(A_{\sdx,\sun})^{-1}\,D_{\sdx,\sun}^{t_{\sun}t_{\sdx}}.
\label{toto}
\ee
 The Yang--Baxter equation for $\cR^{\sun,\sdx}_{11',22'}$ then splits into color and flavor relations. 
The color relations read
\be
&&A_{12}A_{13}A_{23}=A_{23}A_{13}A_{12}\mb{,} D_{12}D_{13}D_{23}=D_{23}D_{13}D_{12};\\ 
&&A_{12}C_{13}C_{23}=C_{23}C_{13}A_{12}\mb{,} D_{12}B_{13}B_{23}=B_{23}B_{13}D_{12}.
\ee
The matrices realizes the well-known reflection Yang--Baxter equations which need not be discussed here.

For the flavor spaces, we get
\be
&&\big(C_{\sun,\sdx}^{t_{\sdx}}\big)^{-1}\,R_{\sun,\sdx}\,B_{\sun,\sdx}^{t_{\sun}}
\big(C_{\sun,\str}^{t_{\str}}\big)^{-1}\,R_{\sun,\str}\,B_{\sun,\str}^{t_{\sun}}
\big(C_{\sdx,\str}^{t_{\str}}\big)^{-1}\,R_{\sdx,\str}\,B_{\sdx,\str}^{t_{\sdx}}
\nonu
&&\qquad=\big(C_{\sdx,\str}^{t_{\str}}\big)^{-1}\,R_{\sdx,\str}\,B_{\sdx,\str}^{t_{\sdx}}
\,\big(C_{\sun,\str}^{t_{\str}}\big)^{-1}\,R_{\sun,\str}\,B_{\sun,\str}^{t_{\sun}}
\,\big(C_{\sun,\sdx}^{t_{\sdx}}\big)^{-1}\,R_{\sun,\sdx}\,B_{\sun,\sdx}^{t_{\sun}}\,.
\ee
We have introduced
\be
R_{\sun,\sdx}=\big(D_{\sun,\sdx}^{t_{\sun}t_{\sdx}}\big)^{-1}\, A_{\sun,\sdx},
\ee
which is unitary, $R_{\sun,\sdx}R_{\sdx,\sun}=\II$ by \eqref{toto}. Remark that the matrices $A$ and $D$ appear only through $R$,
showing a freedom 
\be
 A_{\sun,\sdx}\ \to\  G_{\sun,\sdx} A_{\sun,\sdx} \mb{and}  D_{\sun,\sdx}\ \to\  D_{\sun,\sdx} G_{\sun,\sdx}^{t_{\sun}t_{\sdx}}
 \ee
where $G_{\sun,\sdx}$ is any invertible matrix.

The flavor part has the form of  a Yang--Baxter equation with the following twisted $R$ matrix:
\be
&&\wt R_{\sun,\sdx}\ =\ \big(C_{\sun,\sdx}^{t_{\sdx}}\big)^{-1} R_{\sun,\sdx}C_{\sdx,\sun}^{t_{\sun}}\mb{with}
\wt R_{\sun,\sdx}\,\wt R_{\sun,\str}\,\wt R_{\sdx,\str}=
\wt R_{\sdx,\str}\,\wt R_{\sun,\str}\,\wt R_{\sun,\sdx},
\ee
where we have used \eqref{toto} to get the result. It is easy to see that $\wt R_{\sun,\sdx}$
 is also unitary.
\medskip

 Starting now from any unitary solution to the Yang--Baxter equation $\wt R_{\sun,\sdx}$ and any invertible matrix $F_{\sun,\sdx}$, one reconstructs the matrices $R$, $B$ and $C$ through
\be
B_{\sdx,\sun}=C_{\sun,\sdx}=F_{\sun,\sdx}^{t_{\sdx}}\mb{;}  R_{\sun,\sdx}\ =\ F_{\sun,\sdx}^{-1}\wt R_{\sun,\sdx}F_{\sdx,\sun}\,.
\ee

Remarkably, the splitting of color and flavor spaces induces the usual \textit{reflection} Yang--Baxter equations for the color part, 
whereas we identify a \textit{twisted} Yang--Baxter equation for the flavor part.

\section{Dynamisation}

The next issue is now to define consistent dynamical extensions of the trace R.A. Dynamical extensions of quantum algebra have
a long story going back to the Gervais-Neveu-Felder (also called ``dynamical Yang--Baxter equation'') \cite{GNF}, characterizing the Belavin-Baxter
statistical mechanics $R$-matrix \cite{BB} for IRF models. The general idea is there to introduce a dependance of the $R$-matrix, and the matrix $T$ encapsulating the algebra generators, in so-called ``dynamical'' non-operatorial parameters interpreted as coordinates on the dual 
of some Lie subalgebra of the underlying Lie algebra (or affine algebra) in the quantum structure. In the GNF case, which will be the basis of our derivation here, the subalgebra is the abelian Cartan subalgebra of this Lie/affine algebra (for non-abelian cases see e.g. \cite{Ping}). The $RTT$ relations and associativity
conditions are accordingly modified to yield the GNF-type dynamical Yang--Baxter equations. 


This notion naturally extends to R.A. The first dynamical R.A. was identified as consistency conditions \cite{BPoB} for the boundary
matrix defining open IRF models. It was later studied as ``dynamical boundary algebra'' in \cite{Kor} and \cite{ANR1}. It is identified
with a dynamical twist of a quadruple tensor product of quantum affine $RTT$ algebras \cite{DKM2}.  The second one was identified \cite{ACF}
in the quantum formulation of Ruijsenaar-Schneider models \cite{RS}. It was later studied in \cite{ANR1} and characterized in \cite{AR} as
a deformation of a non-dynamical $RTT$ algebra by a dynamical semi-gauge action. A third one was recently constructed in \cite{ARag} and seems
related to twisted Yangian structures instead of quantum affine algebras.

\subsection{Freidel--Maillet formulation of the dynamical reflection algebras}

In order to define dynamical versions of the trace R.A. we will use the
bivector formulation a la Freidel--Maillet. We must first of all construct such a formulation
for the single-flavor dynamical R.A. described above. To the best of our knowledge this has never been done. The formulation
which we propose follows on these lines:

All three dynamical R.A.'s are represented by similar-looking quadratic exchange algebra relations:
\be\label{DYR-1}
A_{12}(\lambda)K_1(\lambda -\epsilon_R h_2) B_{12}(\lambda)K_2(\lambda +\epsilon_L h_1) = 
K_2(\lambda -\epsilon_R h_1)C_{12}(\lambda)K_1(\lambda +\epsilon_L h_2)D_{12}(\lambda)\
\ee
where $\epsilon_L$ and $\epsilon_R$ are some complex number characterizing 
the different reflection algebras (see below).

The dynamical variables are encapsulated in an $n$ dimensional vector $\lambda$ which will be omitted whenever no ambiguity
of notation arises. It is assumed that the auxiliary vector space on which the structure matrices act is a diagonalizable,
fully reducible module of the Cartan algebra, hence notations such as $K_1(\lambda -\epsilon_R h_2)$ are self
explanatory. They will be used everytime the shift operates along a copy of the dual Cartan algebra \textit{not} acted upon
by the matrix inside which it appears. Shifts along copies of the dual Cartan algebra acted upon by the
matrix itself must be defined in a more specific context. This brings us to define precisely the notions of ``external'' and ``internal''
shifts which will be of use throughout our derivation.

\begin{defi}
We recall the well-known notation of outside action of a shift operator 
(i.e. shift along a Cartan algebra copy $(a)$ not acted upon
by the matrix $M(\lambda)$). It reads:
\be\label{extshift}
e^{(\epsilon h_a\partial)}M_{\dots}(\lambda) e^{(-\epsilon h_a\partial)}= M_{\dots}(\lambda + \epsilon h_a)\,.
\ee
\end{defi}

\begin{defi}
Consider now the problem of inside action. Denote by $M_{... a ...}$ a matrix acted upon by 
$e^{(\epsilon h_{a}\partial)}$. In order to obtain a pure c-number matrix
without explicit difference operators after action of the shifts we must then consider only the following objects:
\be\label{sl}
&&((e^{(\epsilon h_{a}\partial)} M)^{t_a} e^{(-\epsilon h_{a}\partial)})^{t_a}:= M^{sr(a)}\,,
\\
\label{sc}
&&((e^{(\epsilon h_{a}\partial)} (M e^{(-\epsilon h_{a}\partial)})^{t_a}:= M^{sc(a)}\,,
\ee
 where, for conciseness, we have omitted the $\lambda$-dependence in $M$.

One also defines a natural extension of the shift-row procedure to single vector indices:
\be
K_{aa'}^{{\rm s}_{a'}(-\epsilon_L)} := 
\big((e^{(\epsilon h_{a'}\partial)} K_{aa'}\big)^{t_a'} e^{(-\epsilon h_{a'}\partial)})^{t_{a'}}
, \quad a\in \{1,2\}\,.
\ee
\end{defi}

This defines the notations ${\rm sc}_{a}$, ${\rm sr}_{a}$ and ${\rm s}_{a} $. 
We see of course that their application
to a matrix depending on $\lambda$ mean that the matrix
elements are transformed by a shift of their dynamical variables as 
$\lambda_i \rightarrow \lambda_i + \delta_{i,k}$ where $k$ is column (resp. row) index 
in their tensorial factor $(a)$. 
In this way we have taken care of the situation
where shifts occur along copies of the dual Cartan algebra acted upon by the matrix itself.

A number of identities must now be established as interplay between different types of shifts. First of all one has:
\be
&&\label{slsc}
(M^{sr(a)})^{t_a} =  (M^{t_a})^{sc(a)}\,.
\ee
This identity does not follow in a manifest way from \eqref{sl}, \eqref{sc} but must be checked
directly by computing the matrix elements.

Then one can in fact reinterprete outside shift as inside shift of a `completed'' matrix as follows:
\be\label{inout}
M_{\dots}(\lambda + \epsilon h_a) := (M_{...} \otimes \bb I_a)^{sc(a)}:= (M_{...} \otimes \bb I_a)^{sr(a)}\,.
\ee

Shift operations along a space $(a)$ factor out on matrix products only when one of the factor matrices acts diagonally
on this space. One has:
\be\label{diagshc}
(M_{\dots a \dots}D_{\dots a \dots})^{sc(a)} := (M_{\dots a \dots}^{sc(a)} D_{\dots a \dots})^{sc(a)}
\ee
and
\be\label{diagshl}
(D_{\dots a \dots}M_{\dots a \dots})^{sr(a)} := (D_{\dots a \dots}^{sr(a)} M_{\dots a \dots})^{sr(a)}\,,
\ee
where $M$ is any matrix and $D$ is diagonal on space $(a)$. Of course shift-row and shift-column are identical
operations on $D$. Combining \eqref{inout}, \eqref{diagshl} and \eqref{diagshc} yields the identification:
\be\label{fuse1}
(N_{\dots}(\lambda + \epsilon h_a)(M_{\dots a \dots})^{sr(a)} := (N_{\dots} M_{\dots a \dots})^{sr(a)} 
\ee
and dually
\be\label{fuse2}
M_{\dots a \dots}^{sc(a)}N_{\dots}(\lambda + \epsilon h_a) := (M_{\dots a \dots} N_{\dots} )^{sc(a)} \,.
\ee

Let us now consider the particular case of structure matrices.
A key consistency property for dynamical reflection algebras are the zero-weight conditions
of the structure matrices $A,B,C,D$. They must indeed obey:
\be
&&\epsilon_{R}\,{[h^{(1)}+h^{(2)}\,,\,A_{12}]}\ =\  
\epsilon_{L}\,{[h^{(1)}+h^{(2)}\,,\,D_{12}]}\ =\ 0\,,
\label{eq:zerowAD}\\
&&{[\epsilon_{R}\,h^{(1)}-\epsilon_{L}\,h^{(2)}\,,\,C_{12}]}\ =\  
{[\epsilon_{L}\,h^{(1)}-\epsilon_{R}\,h^{(2)}\,,\,B_{12}]}\ =\  0\,.
\label{eq:zerowBC}
\ee
The three dynamical R.A.'s respectively correspond to the choice $\epsilon_R = -1, \epsilon_L  = 1$ (DBA); 
$\epsilon_R = -1, \epsilon_L  = 0$  (so-called semi-dynamical R.A.) ; $\epsilon_R = -1, \epsilon_L  = -1$ (twisted Yangian R.A.).

We now introduce the notion of zero-weight shift which will be in fact particularly relevant to such structure matrices:
\begin{defi}
Consider a matrix $M_{... ab...}$ obeying a zero-weight condition
\be
&&{[\epsilon_{a}\,h^{(a)}+\epsilon_{b}\,h^{(b)}\,,\,M_{... ab...}]}\ =  0\,.
\label{eq:zerowM}
\ee
The exponential of the ``zero-weighted'' shifts $e^{(\epsilon_{a} h_{a} + \epsilon_{b} h_b \partial)}$
does act on the relevant matrix to yield again a pure c-number matrix (no shift term remains) with shifts inside
the matrix elements given in terms of the $sr,sc$ notions. This action yields a crossing shift formula:
\be
\label{crossshift}
e^{(\epsilon_a h_{a}\partial)}{\tilde M_{ab}}e^{(-\epsilon_b h_{b}\partial)}:= e^{(-\epsilon_b h_{b}\partial)}M_{ab}e^{(\epsilon_a h_{a}\partial})\,.
\ee
\end{defi}

In particular from the zero-weight conditions on $A,B,C,D$ one has:
\be\label{conjugA}
&& e^{(\epsilon_R h_{1}\partial)}{\tilde A_{12}}e^{(-\epsilon_R h_{2}\partial)}:= e^{(\epsilon_R h_{2}\partial)}A_{12}e^{(-\epsilon_R h_{1}\partial})\,,
\\
\label{conjugD}
&& e^{(\epsilon_L h_{1}\partial)}{\tilde D_{12}}e^{(-\epsilon_L h_{2}\partial)}:= e^{(\epsilon_L h_{2}\partial)}D_{12}e^{(-\epsilon_L h_{1}\partial})\,,
\\
\label{conjugB}
&& e^{(\epsilon_L h_{1}\partial)}{\tilde B_{12}}e^{(\epsilon_R h_{2}\partial)}:= e^{(\epsilon_R h_{2}\partial)}B_{12}e^{(\epsilon_L h_{1}\partial)}\,,
\\
\label{conjugC}
&& e^{(\epsilon_L h_{2}\partial)}{\tilde C_{12}}e^{(\epsilon_R h_{1}\partial)}:=e^{(\epsilon_R h_{1}\partial)}C_{12}e^{(\epsilon_L h_{2}\partial)}\,.
\ee

An important property of the shift of a product involving a zero-weight matrix is the following:
\begin{prop}\label{fuse}
Given a $A_N$ zero-weight matrix $M_{ab}$ such that $${[h^{(a)}+ h^{(b)}\,,\,M_{ab}]}\ =  0$$ 
(the possible other space labels are omitted) and a matrix $C_{ab}$ (without any weight conditions), then:
\be
(C_{ab} M_{ab})^{sc(a) sc(b)} := C_{ab}^{sc(a) sc(b)} M_{ab}^{sc(a) sc(b)}
\ee
and
\be
(M_{ab} C_{ab})^{sr(a) sr(b)} := M_{ab}^{sr(a) sr(b)} C_{ab}^{sr(a) sr(b)}\,.
\ee
\end{prop}

The proof is by direct computation of the respective matrix elements, using the fact that the set of row and column
indices of a $A_N$ zero-weight matrix are identified. 

We are now able to prove the following:

\begin{thm}
The dynamical R.A. \eqref{DYR-1} is represented in the bivector formalism by the following expression:
\be\label{DYR}
\begin{array}{c}
A_{12}(\lambda-\epsilon_L(h_{2'}+h_{1'}) )\big(B_{1'2}^{t_{1'}}(\lambda-\epsilon_L h_{2'})\big)^{{\rm sr}_{1'}(\epsilon_L)}{\bar K_{11'}}(\lambda-\epsilon_R h_2 -\epsilon_L h_{2'} ){\bar K_{22'}}(\lambda)= 
\\[4mm]
\big(D_{1'2'}^{t_{1'}t_{2'}}(\lambda)\big)^{{\rm sr}_{1'}(\epsilon_L){\rm sr}_{2'}(\epsilon_L)}
\big(C_{12'}^{t_{2'}}(\lambda-\epsilon_L h_{1'})\big)^{{\rm sr}_{2'}(\epsilon_L)}
{\bar K_{22'}}(\lambda-\epsilon_R h_1-\epsilon_L h_{1'}){\bar K_{11'}}(\lambda)\,.
\end{array}
\ee
\end{thm}

The proof is a long and delicate  (but rather straightforward) computation. It requires first of all to rewrite \eqref{DYR-1}
using explicit shift operators of the general form $e^{(-\epsilon_{L/R} h_{2/1}\partial)}$.

$$
\begin{array}{c}
A_{12}e^{(-\epsilon_R h_{2}\partial)}K_{11'}^{t_{1'}}e^{(\epsilon_R h_{2}\partial)}B_{1'2}e^{(\epsilon_L h_{1'}\partial)}K_{22'}^{t_{2'}}=
\\[4mm]e^{(-\epsilon_R h_{1}\partial)}K_{22'}^{t_{2'}}e^{(\epsilon_R h_{1}\partial)}C_{12'}e^{(\epsilon_L h_{2'}\partial)}K_{11'}^{t_{1'}}e^{(-\epsilon_L h_{2'}\partial)}D_{1'2'}e^{(\epsilon_L  h_{1'}\partial)}\,.
\end{array}
$$
Partial transposition with respect to space indices $1'$ and $2'$ redefines as before $K$ as bivectors instead of matrices. Using the cross-shift properties \eqref{conjugD}, \eqref{conjugB}, \eqref{conjugC} and \eqref{conjugA} allows
to rewrite the previous equality as:
$$
\begin{array}{c}
A_{12}e^{(-\epsilon_R h_{2}\partial)}({\tilde B_{1'2}}^{t_{1'}})^{\rm{sc}_{1'}(\epsilon_L)}e^{(\epsilon_L h_{1'}\partial)}K_{11'}^{\rm{s}_{1'}(-\epsilon_L)} e^{(\epsilon_R h_{2}\partial)}e^{(\epsilon_L h_{2'}\partial)}K_{22'}^{{\rm s}_{1'}(-\epsilon_L)}=
\\[1ex]
(({\tilde C}_{12'}({\tilde D_{1'2'}}(\epsilon_R  h_{1}))^{\rm{sc}_{1'}(\epsilon_L})^{t_{2'}})^{\rm{sc}_{2'}(\epsilon_L)}e^{(\epsilon_L h_{2'}\partial)}K_{22'}^{{\rm s}_{2'}(-\epsilon_L)}
e^{(\epsilon_R h_{1}\partial)}e^{(\epsilon_L h_{1'}\partial)}K_{11'}^{{\rm s}_{1'}(-\epsilon_L)}.
\end{array}
$$
In the equation above and in the following, only shifts in the dynamical parameter are indicated, and for instance $D_{1'2'}(\epsilon_R  h_{1})$ stands for $D_{1'2'}(\lambda+\epsilon_R  h_{1})$.

Pushing the shift operators $ e^{(\epsilon_R h_{2}\partial)}e^{(\epsilon_L h_{2'}\partial)} $ and
$ e^{(\epsilon_R h_{1}\partial)}e^{(\epsilon_L h_{1'}\partial)} $ to the left, using the definition of the 
internal shifts allows for undoing the cross-shift of $A,B,C,D$ to yield:
\be\label{DYRint}
\begin{array}{c}
A_{12}(-\epsilon_L(h_{2'}+h_{1'}) )(B_{1'2}^{t_{1'}}(-\epsilon_L h_{2'}))^{{\rm sr}_{1'}(\epsilon_L)}{\bar K_{11'}}( -\epsilon_R h_2 -\epsilon_L h_{2'} ){\bar K_{22'}}= 
\\[4mm]
[(C_{12'}(-\epsilon_L h_{1'}) (D_{1'2'}^{{\rm sc}_{1'}(-\epsilon_L)})^{t_{1'}})^{{\rm sr}_{2'}(-\epsilon_L)}]^{t_{2'}}{\bar K_{22'}}(-\epsilon_R h_1-\epsilon_L h_{1'}){\bar K_{11'}}\,.
\end{array}
\ee

The structure matrices on the l.h.s. are now decoupled. To achieve a similar decoupling of $[(C_{12'}(-\epsilon_L h_{1'}) (D_{1'2'}^{{\rm sc}_{1'}(-\epsilon_L)})^{t_{1'}})^{{\rm sr}_{2'}(-\epsilon_L)}]^{t_{2'}} $ on the r.h.s. we essentially use eqs \eqref{slsc}, \eqref{inout},
Proposition \ref{fuse} and eq \eqref{fuse2} to finally yield the decoupled terms: 

$(D_{1'2'}^{t_{1'}t_{2'}})^{{\rm sr}_{1'}(\epsilon_L){\rm sr}_{2'}(\epsilon_L)}(C_{12'}^{t_{2'}}(-\epsilon_L h_{1'}))^{{\rm sr}_{2'}(\epsilon_L)} $.

This representation of the dynamical reflection algebras allows to identify some key features of dynamical reflection algebras which shall
be crucial guidelines in our conjectured formulation of a dynamical quantum reflection algebra. The bivector representation is indeed essential to identify these features and represent probably a deeper formulation of reflection algebras in general.

{\bf Criterion 1} The shifts separate into shifts labeled by the tensorial factors generated by the original vector-type indices 
in $K$ (row indices, unprimed) weighted by $-\epsilon_R$; and shifts labeled by the tensorial factors corresponding to the 
original covector-type indices in $K$ (column indices, primed) weighted by $-\epsilon_L$.

{\bf Criterion 2} The structure matrices $A,B,C,D$ are zero-weighted according to the nature of their tensorial labels with the
proviso that the zero-weight conditions be written for the {\it partially transposed} matrices such as occur 
in the bivector formulation.

{\bf Criterion 3} All four structure matrices are shifted along both primed-labeled directions, weighted by $-\epsilon_L$. Depending whether
these labels occur or not in the matrix the shifts are inside shifts (resp. outside).

{\bf Criterion 4} The $K$ matrices are shifted along three directions: the two respective outside shifts and the inside prime (transposed covector) shift occur with their respective consistent weights of Criterion 1.

\subsection{Conjectural dynamical quantum reflection trace algebra}

The quantum reflection trace algebra structure essentially differs from the QRA by the occurence of the flavor vector
index in $K$ and the pair of corresponding extra auxiliary spaces in $A,B,C,D$. Our key hypothesis is to treat this extra vector
index in $K$ as a supplementary `true'' vector index (unprimed). The dynamization will now also contain a deformation parametrized
by coordinates on the dual of the abelian Cartan subalgebra of the new flavor Lie algebra, a priori here $A_{m-1}$. Once again one assumes
that the suppplementary flavor vector spaces are diagonalizable fully reducible modules of this flavor Cartan agebra.

Accordingly we now introduce a third weight $\epsilon_f$
for the associated shifts (criterion 1); complement the zero-weight conditions on $A,B,C,D$ by this extra weight and the extra
generators of the abelian Cartan subalgebra of the new flavor Lie algebra (criterion 2); do not modify
the shift structure on the matrices $A,B,C,D$ themselves (since this extra vector index is a `true'' index, not a transposed covector) (criterion 3); shift the $K$ matrices additionally along their \textit{outside} flavor space (criterion 4).

We thus propose the following form where the flavor space labels have been omitted for the sake of simplicity, and as in \eqref{FMform}
the transpositions are defined as $T_{1'}\equiv t_{1'}t_{\sun}$ and $T_{2'}\equiv t_{2'}t_{\sdx}$. 
\be\label{DTrAl}
\begin{array}{c}
A_{12}(\lambda-\epsilon_L(h_{2'}+h_{1'}) )\big(B_{1'2}^{T_{1'}}(\lambda-\epsilon_L h_{2'})\big)^{{\rm sr}_{1'}(\epsilon_L)}{\bar K_{11'}}(\lambda -\epsilon_R h_2 -\epsilon_L h_{2'} - \epsilon_f h_{\sdx} ){\bar K_{22'}}(\lambda)= 
\\[4mm]
\big(D_{1'2'}^{T_{1'}T_{2'}}(\lambda)\big)^{{\rm sr}_{1'}(\epsilon_L){\rm sr}_{2'}(\epsilon_L)}\big(C_{12'}^{T_{2'}}(\lambda-\epsilon_L h_{1'})\big)^{{\rm sr}_{2'}(\epsilon_L)}{\bar K_{22'}}(\lambda-\epsilon_R h_1-\epsilon_L h_{1'} - \epsilon_f h_{\sun}){\bar K_{11'}}(\lambda)\,.
\end{array}
\ee

Although the transpositions $T$ contain transpositions of the {\it flavor}-labeled components in the structure matrices $A,B,C,D$
we have decided here to formulate and apply Criterion 3 so as to not shift the corresponding matrices along these {\it flavor} 
transposed space, since we interpret the shifts of $B,C,D$ in \eqref{DYRint} to only operate in directions corresponding to
transposed covector indices in $K$. There is however a possible ambiguity since one could also interprete more broadly these shifts
as occuring along all directions characterized by a transposed label. In this case one should reinterpret ${\rm sr}_{1'}, {\rm sr}_{2'} $
as ${\rm sr}_{1'} + {\rm sr}_{\sun}, {\rm sr}_{2'}+ {\rm sr}_{\sdx}$ in $B,C,D$ and $\bar K$ wherever they occur in \eqref{DTrAl}. Lifting the ambiguity requires an in-depth study of these structures which we shall leave for a later investigation.

\section{Conclusion}

We have established a consistent form for the quantum trace reflection algebra, and conjectured a form for an abelian dynamical
deformation, on the lines defined by the flavorless dynamical reflection algebras. We wish to emphasize once again that the rewriting of
reflection algebras (particularly dynamical) in a bivector formalism plays a crucial role in that it has allowed us to extract what appear to be the key features of this type of dynamical deformation. It suggests that the bivector Freidel--Maillet formulation is possibly the
relevant frame to understand in depth the quantum reflection algebra (a point always defended by the authors of \cite{FM}
and consistent with the construction of reflection algebras by twisting a quadrupled algebra in \cite{DKM}, \cite{DKM2}).

It is worth stressing here that our approach to the quantization problem of trace-Poisson brackets and the very formulation of this problem has no immediate straightforward  application to the natural question of a proper deformational quantization of the {\it Van den Bergh double Poisson structure}
itself as it was addressed by D. Calaque \footnote{ See
http://mathoverflow.net/questions/29543/what-is-a-double-star-product}: "Is there a notion of "quantization", or "double star-product" for double Poisson algebras, so that it would induce genuine star-products quantizing the above mentioned (i.e. coordinate rings of the representation moduli spaces) Poisson varieties ?" Or, in other words "what kind of algebraic structure on an algebra $A$ ensures that one will get star-products on $\rm{Rep}_N(A)$?". Anyway, we hope that the scheme proposed here will help to clarify some aspects of the problem.

Let us also stress that we have not adressed here the case of linear double brackets and their subsequent trace-Poisson brackets. Their structure is described (\ref{notation1})  by a mixed object $B$ carrying both matrix and vector flavor indices. By contrast the quadratic double brackets yield quadratic trace-Poisson ``structure constants''  with pure matrix flavor indices, allowing for a quantization and dynamization on more familiar lines as Yang-Baxter $R$-type matrices . It is not clear at this stage how to adress the linear case and we shall postpone its discussion for the time being.

Besides this issue several deep questions are now opened following our derivations. A few have already been mentioned: how about the parametric AYB algebras ?
This of course immediately begs the question of elliptic-type deformations (i.e. along the $d$ generator of an underlying
\textit{affine} algebra. In addition
we should ask the question of a co-algebra or (more probably) co-ideal structure in the non-dynamical case, in relation with the
quasi-Hopf structure which it seems to exhibit. Finally in the dynamical case one must now adress the issues both of consistency and relevance (physical and mathematical) of this conjectured structure (whichever version of Criterion 3 turns out to be consistent).
The existence of a full rewriting procedure, reversing the previous bivectorializations, of this structure in terms of $K$ 
{\it matrices} as an explicit reflection
formula mimicking \eqref{FMform} with suitable internal and external shifts, may provide a good consistency criterion.

\subsection*{Acknowledgements}

This work was sponsored by CNRS; Universit\'e de Cergy-Pontoise; Universit\'e de Savoie; 
and ANR Project DIADEMS (Programme Blanc ANR SIMI1 2010-BLAN-0120-02). J.A. and V.R. wish to thank
LAPTh for their kind hosting. The work of V.R. was partially supported by RFBR grant 12-01-00525. He is thankful
to A. Odesskii and V. Sokolov for  discussions and collaboration on related subjects.

\end{document}